\documentclass[12pt]{article}
\usepackage[dvips]{graphicx} 
\begin{document}

\begin{center} {\large \bf
  Focus points and the Lightest Higgs Boson Mass in the Minimal Supersymmetric Standard Model}\bigskip\\
\end{center}

\begin{center}  {\large  E. Jur\v{c}i\v{s}inov\'a$^{1,2}$ and M. Jur\v{c}i\v{s}in$^{1,3}$
} \vspace*{1.5mm}
\end{center}

\begin{center}
{\it $^1$ Institute of Experimental Physics, Slovak Academy of
  Sciences, \\ Watsonova 47, 04353 Ko\v{s}ice, Slovakia} \\
{\it $^2$ Laboratory of Information Technologies, Joint Institute
for Nuclear Research,\\ 141 980 Dubna, Moscow Region, Russian Federation} \\
 {\it $^3$ Laboratory of Theoretical Physics, Joint Institute for
  Nuclear Research, \\ 141 980 Dubna, Moscow Region, Russian Federation }

 e-mail: jurcisin@thsun1.jinr.ru, eva.jurcisinova@post.sk
\end{center}

\begin{abstract}
 We investigate focus points of the
renormalization group equations of the Minimal Supersymmetric
Standard Model. We show that within this model the up- and down-type
Higgs mass soft supersymmetry breaking parameters have focus point
behavior at the electroweak scale simultaneously when appropriate
conditions are fulfilled. The focus point scenario is holding for
large $\tan \beta$. This two focus point scenario allows to fix the
pole top-quark mass which is within the experimentally allowed
interval. The main goal of the present paper is the investigation of
the influence of the existence of focus points on the determination
of the mass of the lightest Higgs boson.
\end{abstract}

\section{Introduction}

\noindent During the last two decades the idea of supersymmetry was
the most promising assumption at high energies. The simplest
supersymmetric extension of the Standard Model of elementary
particles physics is Minimal Supersymmetric Standard Model (MSSM)
(see for example \cite{mssm,kazakov}). When working within the MSSM,
one encounters large parameter freedom which is mainly due to the
so-called soft SUSY breaking terms \cite{mssm,kazakov}. At the same
time, a large number of free parameters decrease the predictive
power of a theory. The simplest way to reduce this freedom is to
make some assumptions at a high energy scale (for example, at the
Grand Unification (GUT) scale or at the Planck scale). Then,
treating the MSSM parameters as running variables and using the
renormalization group equations (RGEs), one can drive their values
at a low-energy scale. The most common assumption is the so-called
universality of the soft supersymmetry breaking terms, which means
an equality of some parameters at a high energy scale. Adopting the
universality, one reduces the parameter space to a five-dimensional
one given by: a common scalar mass $m_0$, a common gaugino mass
$m_{1/2}$, a common trilinear scalar coupling $A$, a supersymmetric
Higgs-mixing mass parameter $\mu$, and a bilinear Higgs coupling
$B$. The last two parameters can be eliminated in favor of the
electroweak symmetry breaking scale, $v^2=v_1^2+v_2^2= (174.1
GeV)^2$, and the Higgs fields vevs ratio $\tan\beta=v_2/v_1$, when
using minimization conditions of the Higgs potential.

Further reduction of the parameter space of the MSSM can be achieved
using the concept of the so-called infrared quasi fixed points
(IRQFPs) \cite{hill}. Over the last ten years a great interest was
paid to the phenomenological consequences of the IRQFP behavior of
the corresponding system of the RGEs within the MSSM
\cite{irqfp,ja2,ja3} as well as within the Next to the MSSM (NMSSM)
\cite{nmssm}.

However, the IRQFP scenario can be directly used and works properly
only in the case of small $\tan\beta \sim 1$ regime (see, e.g.,
discussion in Ref.\,\cite{ja3}). On the other hand, the small
$\tan\beta$ scenario is already excluded by the recent experimental
data \cite{igo}, therefore the moderate and large $\tan\beta$
regimes come to be investigated from the phenomenological point of
view.

Recently the idea of so-called focus point was used in the
investigation of the system of RGEs of the MSSM \cite{feng}. This
means that the RG trajectories of some parameter of the model may
meet at a "focus point", where their values are independent of their
ultraviolet boundary values. In the Refs.\,\cite{feng} the focus
point behaviour of the up-type Higgs mass parameter was
investigated. The different aspects of this idea was then discussed
in several papers \cite{focuss}.
 In present paper we adopt
the strategy based on the focus point behavior of the RGEs in the
analysis of the mass of the lightest Higgs boson. It is found that
within the MSSM the up- and also down-type Higgs mass soft
parameters have focus point behavior at the electroweak scale
simultaneously when appropriate conditions are fulfilled. This leads
to the determination of the top-quark mass. Thus, if the focus point
scenario works, the lightest Higgs boson mass is determined more
precisely. As we shall see the mass of the lightest Higgs boson
determined by this method is allowed by recent experiment
\cite{igo}.

\section{Focus points of the RGEs}

\noindent In this section, we discuss the phenomenon of the focus
points in the RG evolution of supersymmetry breaking parameters. A
detailed mathematical treatment of such behavior can be found in
Refs.\,\cite{feng}.

We shall use the following notations \cite{ja2}: for Yukawa
coupling constants $h_i, i=t,b,\tau$ we also use expresion
$\rho_i=Y_i/\tilde\alpha$ where $Y_i=h_i^2/(4\pi)^2$ and
$\tilde\alpha_3=\alpha_3/(4 \pi)=g_3^2/(4 \pi)^2$ is strong
coupling constant ($t,b,\tau$ correspond to top quark, botom
quark, and $\tau$ lepton). For trilinear scalar coupling $A_i$ we
also use definition $\rho_{A_i}=A_i/M_3$, where $M_3$ represents
the gluino mass. $\alpha_0=\alpha_{30}$ is the unified coupling
constant at the GUT scale.

\begin{figure}[t]
     \vspace{-1cm}
       \begin{flushleft}
       \leavevmode
       \epsfxsize=8.5cm
       \epsffile{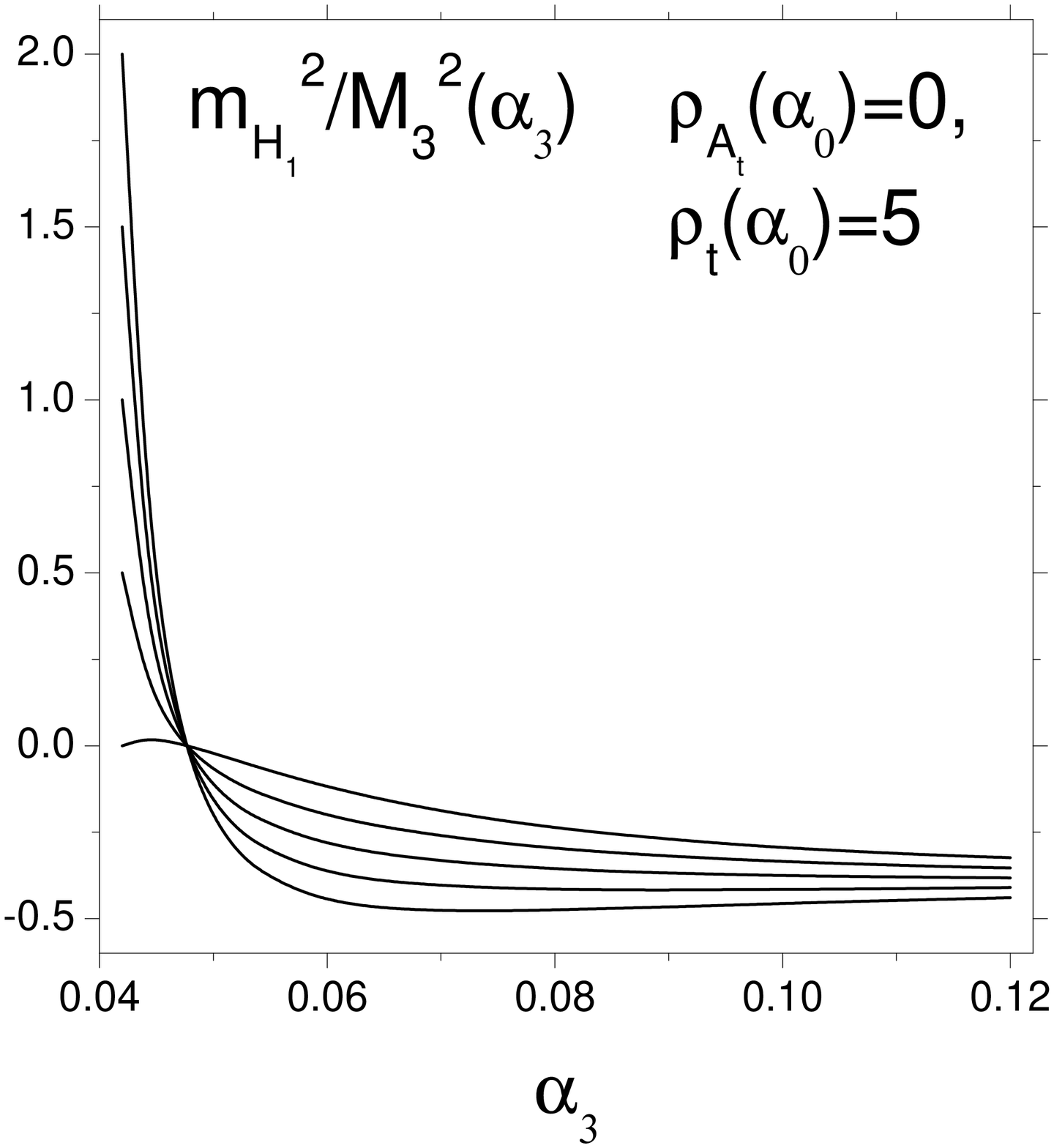}
   \end{flushleft}
     \vspace{-13.2cm}
   \begin{flushright}
       \leavevmode
       \epsfxsize=8.5cm
       \epsffile{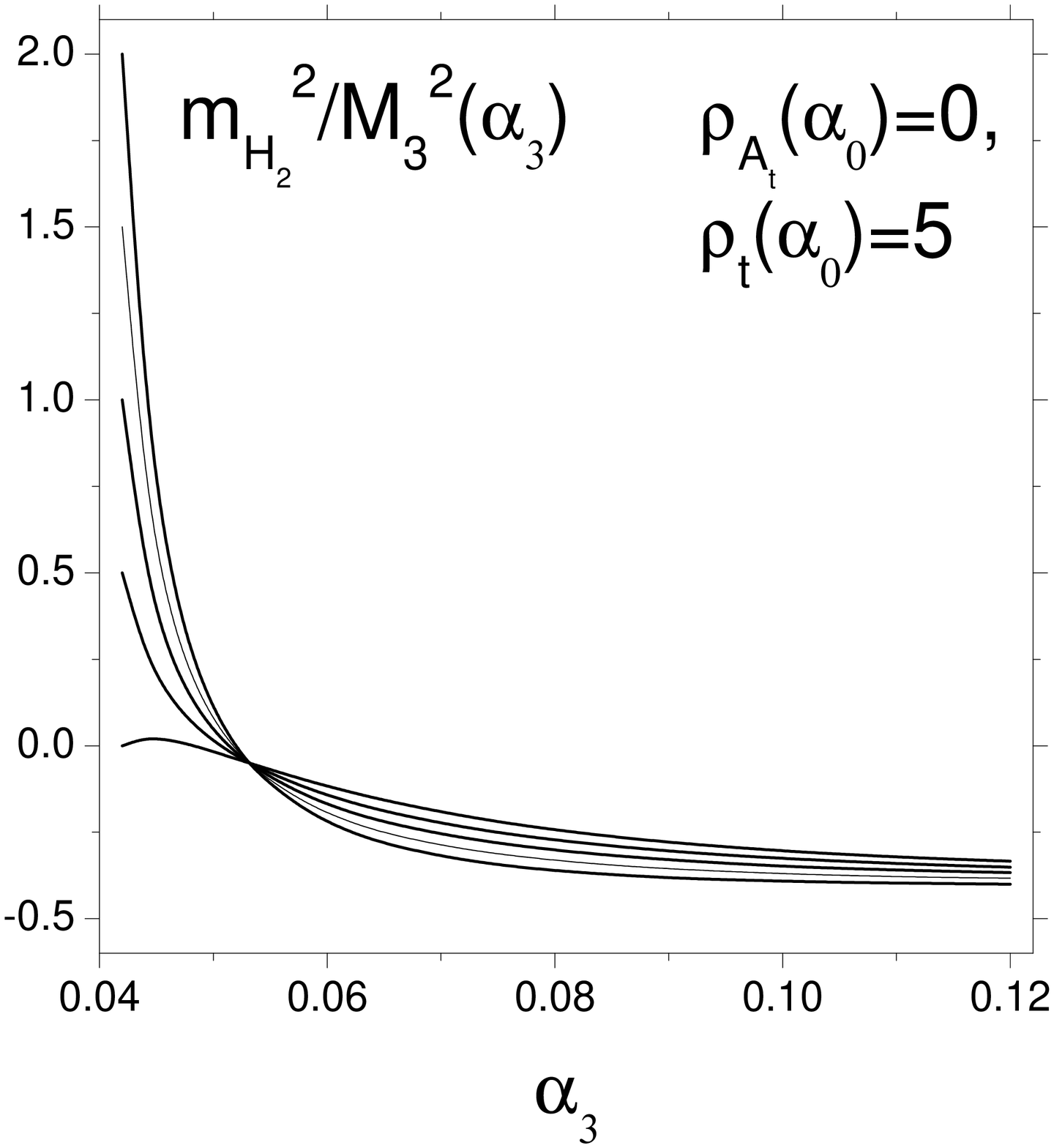}
   \end{flushright}
\vspace{-1.5cm} \caption{The focus points behavior for soft mass
parameters $m_{H_1}^2$ and $m_{H_2}^2$ for the concrete universal
values of Yukawa coupling constants and trilinear scalar couplings
(that means
$\rho_t(\alpha_0)=\rho_b(\alpha_0)=\rho_{\tau}(\alpha_0)$ and
$\rho_{A_t}(\alpha_0)=\rho_{A_b}(\alpha_0)=\rho_{A_{\tau}}(\alpha_0)$).
The value $\alpha_3 \approx 0.042$ corresponds to the GUT scale,
and the value $\alpha_3\approx 0.12$ corresponds to the
electroweak scale. \label{fig1}}
\end{figure}

By complete analysis of the system of the one-loop RGEs in the MSSM
one can find that focus point behavior is connected with the Higgs
mass soft parameters $m_{H_1}^2$ and $m_{H_2}^2$ (the explicit form
of all one-loop RGEs in the MSSM can be found, e.g., in
Ref.\,\cite{kazakov}). In Fig.\,\ref{fig1} is present their running
for different initial values at the GUT scale ($M_{GUT}=2 \cdot
10^{16}$) and where the focus points are shown explicitly.

\input epsf

\begin{figure}[t]
     \vspace{-1cm}
       \begin{flushleft}
       \leavevmode
       \epsfxsize=8.5cm
       \epsffile{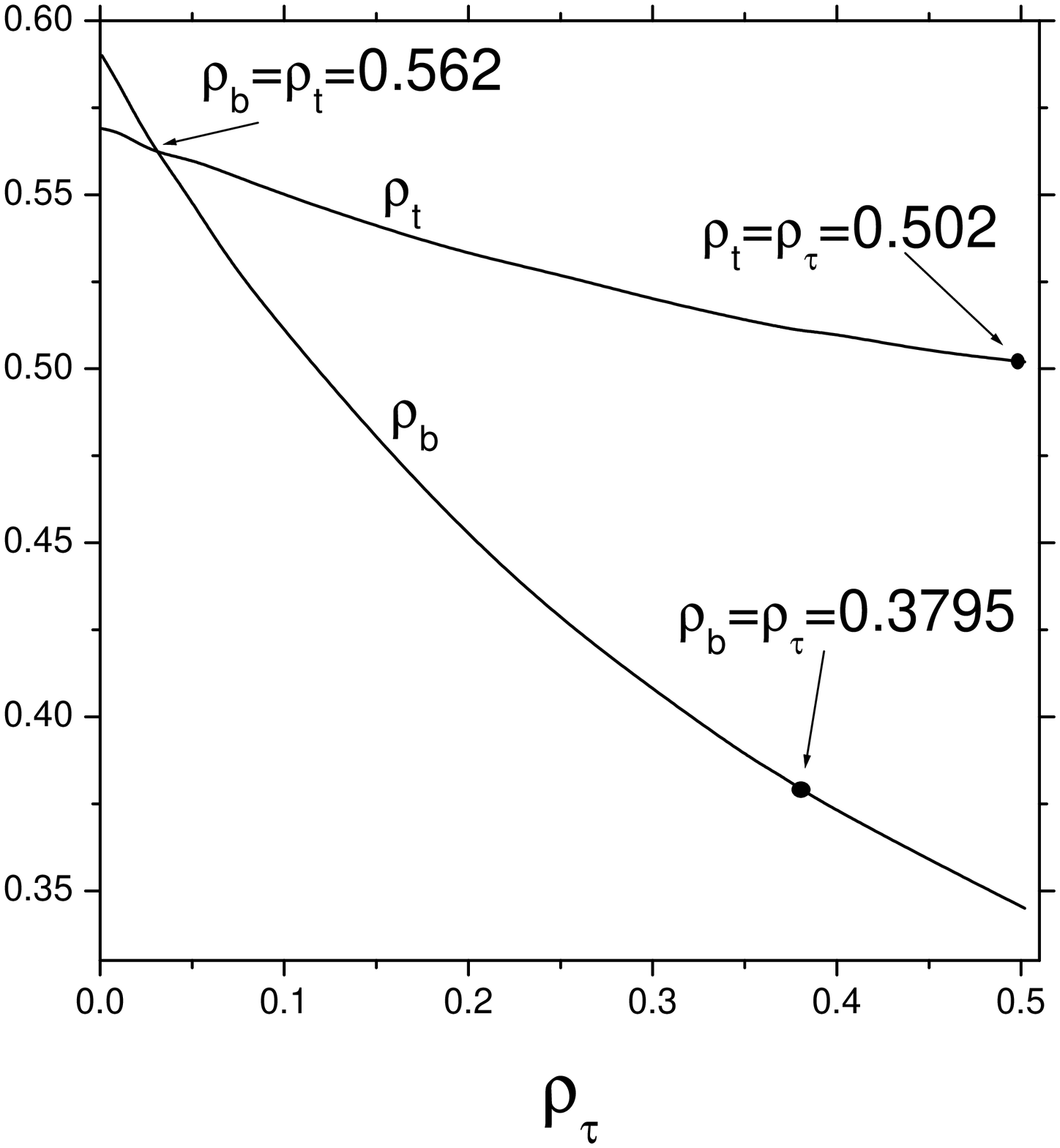}
   \end{flushleft}
     \vspace{-13.2cm}
   \begin{flushright}
       \leavevmode
       \epsfxsize=8.5cm
       \epsffile{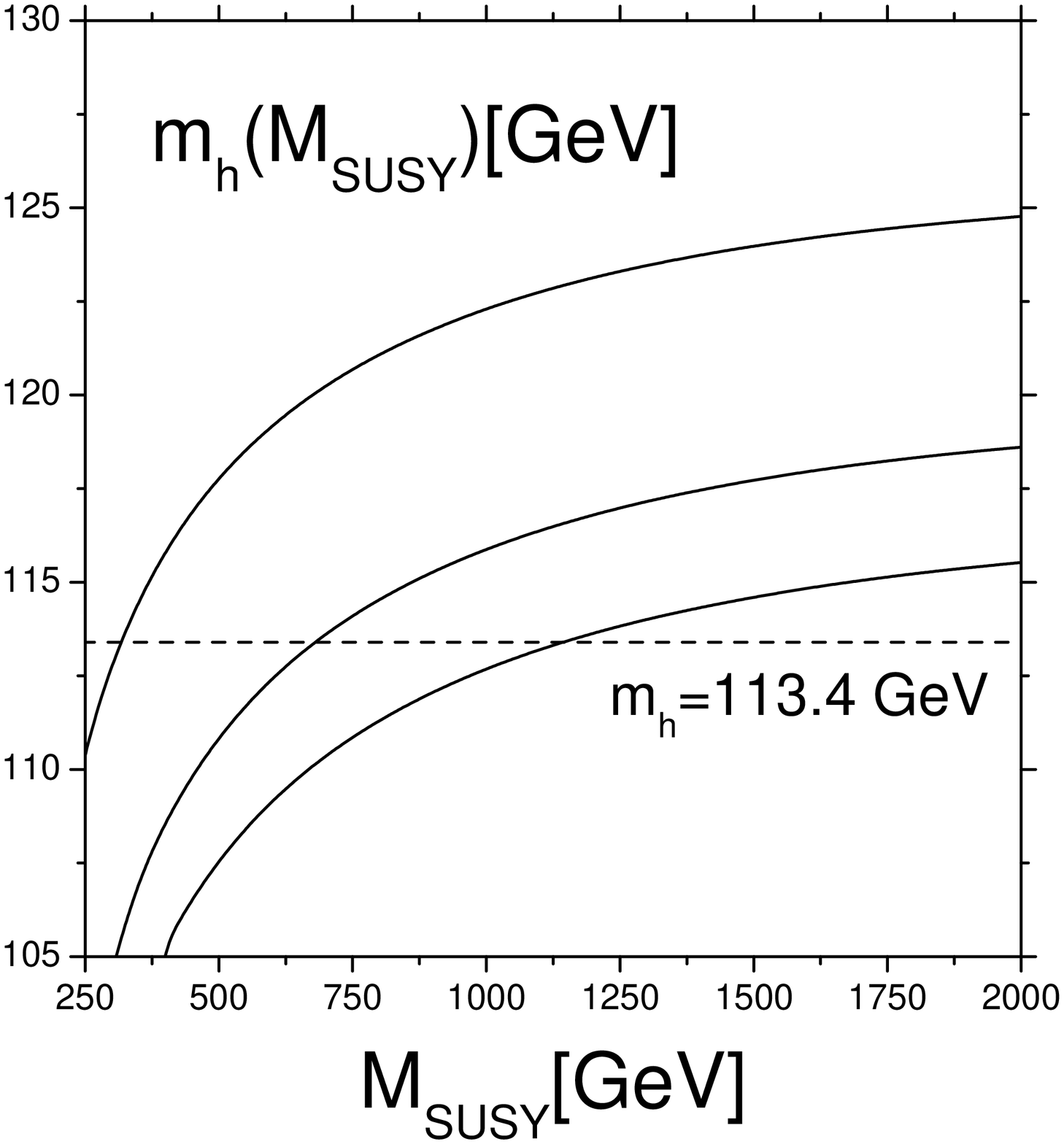}
   \end{flushright}
\vspace{-1.5cm} \caption{(Left) The values of the Yukawa couplings
for which the up- and down-quark Higgs mass soft parameters have
focus points at the electroweak scale simultaneously. The marked
points and the crossing of the lines correspond to the situation
when two initial values of the Yukawa couplings at the GUT scale
are the same. (Right) The mass of the lightest Higgs boson in the
focus point scenario with universal assumption about soft SUSY
breaking parameters. The central line is related to the central
values of the parameters. The upper and lower  lines correspond to
the maximal allowed deviations of the soft parameters. The dash
line represents the experimental restriction of the Higgs boson
mass \cite{igo}.  \label{fig2}}
\end{figure}

Our aim is to analyze if it is possible to have both focus points of
the RGEs for up- and down-type Higgs mass parameter at the
electroweak scale simultaneously.  Using numerical calculations it
is shown that such situation is possible if the Yukawa coupling
constants $Y_i$ have appropriate initial values at the GUT scale. In
our investigation we suppose the universal behaviour of the soft
SUSY breaking parameters at the GUT scale. Thus, the parameter space
of the model is almost completely reduced. In Fig.\,\ref{fig2} are
present the simultaneous values of the Yukawa coupling constants at
the GUT scale which leads to the focus points for soft Higgs mass
parameters at the electroweak scale. It can be shown that the
influence of the initial values of the gaugino mass soft parameters
and the trilinear scalar couplings on the position of the focus
points is negligible. On the other hand, as we shall see in the next
section, their values will have some non-negligible impact on the
determination of the lightest Higgs boson mass.

\input epsf

\section{The focus points and the lightest Higgs boson mass}

\noindent In this section we concentrate our attention to the
determination of the lightest Higgs boson mass based on the focus
point scenario discussed in previous section.

We begin with the description of our strategy. As input parameters
we take the known values of the top-quark, bottom-quark and
$\tau$-lepton pole masses ($m_t^{pole}=(172.7\pm 2.9)$ GeV
\cite{rev0}, $m_b^{pole}=(4.94 \pm 0.15)$ GeV \cite{davies},
$m_{\tau}^{pole}=(1.7771\pm 0.0005)$ GeV \cite{rev1}), the
experimental values of gauge couplings \cite{rev} $\alpha_3=0.120
\pm 0.005, \ \alpha_2=0.034,\ \alpha_1=0.017$ and  the sum of the
Higgs vev's squared $v^2 = v_1^2+v_2^2 \approx $174.1 GeV$^2$. Using
the focus point scenario which determine the values of the Yukawa
coupling constants at the GUT scale, we proceed to the determination
of the $\tan\beta$ and top and bottom quark masses which are related
by well-known relations
\begin{eqnarray}
m_t&=&h_t v \sin\beta\, \label{mt} \\
m_b&=&h_b v \cos\beta\, \label{mb}\\
m_{\tau}&=&h_{\tau} v \cos\beta\, \label{mtau}
\end{eqnarray}
where $m_i,\,\, i=t,b,\tau$ are running quark and lepton masses. The
aim is to find such values of $\tan\beta$ and top-quark,
bottom-quark and $\tau$-lepton masses to be inside the intervals
allowed by experiment and, at the same time, to fulfill
Eqs.\,(\ref{mt}-\ref{mtau}) in the framework of our focus point
scenario. It can be shown that this problem has solutions with
$\tan\beta \simeq 60$. The most problematic in the process of
calculation is to obtain the proper mass of the $\tau$-lepton. In
our calculations we determine the top-quark and bottom-quark running
masses from the corresponding pole masses taking into account QCD
and SUSY corrections \cite{mtop1,mtop2} (for details see
Ref.\,\cite{ja2})
\begin{equation}
m_i(m_i)=\frac{m_i^{pole}}{1+ \left(\frac{\Delta
m_i}{m_i}\right)_{QCD} + \left(\frac{\Delta
m_i}{m_i}\right)_{SUSY}},\,\,\,i=t,b. \label{pole}
\end{equation}
The results depend on the sign of the $\mu$ parameter which enters
the mixing terms in the stop sector. In what follows, we shall
analyze only the case $\mu>0$. The case $\mu<0$ is similar with the
almost the same results and conclusions for the mass of the lightest
Higgs boson mass. The value of the Higgs mixing parameters $\mu$ can
be found from the requirement of radiative electroweak symmetry
breaking and can be determined from the Higgs potential mimimization
condition. The one-loop minimization condition reads
\begin{equation}
\frac{M_Z^2}{2}+\mu^2=\frac{m_{H_1}^2+\Sigma_1-
(m_{H_2}^2+\Sigma_2) \tan^2\beta}{\tan^2\beta-1}\,, \label{MZC}
\end{equation}
where $\Sigma_1$ and $\Sigma_2$ are the one-loop corrections
\cite{Gl}, $M_Z$ is the $Z$-boson mass.

In the MSSM, the Higgs sector consists of five physical states: two
neutral CP-even scalars $h$ and $H$, one neutral CP-odd scalar $A$,
and two charged Higgs scalars $H^{\pm}$. In what follows we shall
concentrate on the mass of the lightest Higgs boson $h$. At the tree
level, the mass of $h$ is smaller than the mass of $Z$-boson, $M_Z$,
but the loop corrections increase it. In general, the mass matrix
for the CP-even neutral  Higgs bosons looks like
\begin{eqnarray}
{\mathcal M} \hspace{-0.3cm}&=& \hspace{-0.4cm}
\left[\left(\begin{array}{cc}\tan\beta & -1\\ -1 & \cot\beta
\end{array} \right)m^2_A
+  \left(\begin{array}{cc}\cot\beta & -1\\ -1 & \tan\beta
\end{array} \right)M^2_Z\right]\cos\beta\sin\beta
 +  \left(\begin{array}{cc} \Delta_{11} & \Delta_{12}\\
\Delta_{12} & \Delta_{22}
\end{array} \right)  \label{h}
\end{eqnarray}
where $m_A$ is the mass of the CP-odd Higgs boson and $\Delta's$
are the radiative corrections~\cite{cor}.

To find the Higgs boson mass one has to diagonalize the mass matrix
(\ref{h}). In the present paper we use the concept of focus points
and universality conditions for soft SUSY breaking parameters. The
Yukawa couplings are determined by focus points to be at the
electoweak scale and soft parameters are inside the
following intervals at the GUT scale: $A_0=A_{0i}/M_{03}\in \\
<-3,3>, i=t,b,\tau$, $m_0^2=m_{0i}^2/M_{03}^2 \in <0.25,4>,\,\,
i=Q, U, D, H_1, H_2$, where $Q$ refers to the third generation
squark doublet, $U$ to the stop singlet and $D$ to the sbottom
singlet.

In Fig.\,\ref{fig2}, the dependence of the mass of the lightest
Higgs boson $m_h$ on the geometric mean of the stop masses
$\sqrt{\tilde m_{t_1} \tilde m_{t_2}}$ (which is often identified
with the supersymmetry breaking scale $M_{SUSY}$) is shown for the
case when bottom-quark and $\tau$-lepton Yukawa couplings at the GUT
scale are equal. As the central values of the parameters we take:
$A_{0t}/M_{03}=0$, $m_{0}^2/M_{03}^2=1$. Taking into account
corresponding deviations from the central values, the mass of the
lightest Higgs boson at a typical scale $M_{SUSY}=1$ TeV ($\mu>0$)
is
\begin{equation}
m_h=115.9 \ ^{\displaystyle +6.4}_{\displaystyle -3.2}\ \pm0.4 \
\mbox{GeV}, \ \ \ \mbox{ for} \ M_{SUSY}=1 \ \mbox{TeV}.
\label{mass2}
\end{equation}
The first uncertainty is given by the deviations from central values
of the soft breaking parameters, and the second one by uncertainty
in the strong coupling constant. Our main result is that in this
approach the dependence on the top-quark mass disappeared
completely. For the central value of the presented result
(\ref{mass2}) we find $m_t^{pole}=174.2$ GeV and $\tan\beta=58$. It
is important that the top-quark mass obtained by the focus point
scenario is inside of experimentally allowed interval \cite{rev0}.
One can see that in the focus point scenario the mass of the
lightest Higgs boson typically belongs to the interval $<113, 122>$
GeV. If we compare our results to the experimental restriction to
the mass of the lightest Higgs boson \cite{igo} $m_h>113.4$, one can
conclude that there still exists a little space to find Higgs boson
related to the MSSM.

\section{Conclusions}

\noindent We have analyzed the behavior of the system of RGEs in the
MSSM from the point of view of focus point behavior. We have found
that such type of behavior is typical for up- and down-quark Higgs
mass soft parameters and that it is possible to have both focus
points at the electroweak scale simultaneously which leads to the
large $\tan\beta$ regime. This non-trivial fact results in further
reduction of the parameter space of the model. The most important
conclusion is that the uncertainty in the top-quark mass disappeared
completely. The mass of the lightest Higgs boson is obtained and it
is not excluded by the experimental restrictions.

\vspace{0.8cm} \noindent {\bf Acknowledgments} \vspace{0.2cm}

\noindent M.J. gratefully acknowledges the hospitality of the
Theoretical division of the Physical Department of CERN. The work
was supported by grant RFFI-RFBR\,05-02-17603.


\begin{thebibliography}{99}
\bibitem{mssm}
H.P.~Nilles, Phys. Rep. {\bf 110} (1984) 1.
\bibitem{kazakov}
D.I.~Kazakov, Survey High Energy Phys. {\bf 10} (1997) 153.
\bibitem{hill}
C.T.~Hill, Phys. Rev. {\bf D24} (1981) 691; \\ C.T.~Hill,
C.N.~Leung, and S.~Rao, Nucl. Phys. {\bf B262} (1985) 517.
\bibitem{irqfp}
M.~Carena et al., Nucl. Phys. {\bf B419} (1994) 213;\\ W.~Bardeen
et al., Phys. Lett. {\bf B320} (1994) 110;\\ M.~Carena and
C.E.M.~Wagner, Nucl. Phys. {\bf B452} (1995) 45;\\ P.~Langacker
and N.~Polonsky, Phys. Rev. {\bf D50} (1994) 2199;\\ M.~Lanzagorta
and G.G.~ Ross, Phys. Lett. {\bf B349}(1995) 319; \\I.~Jack,
D.R.T.~Jones and K.L.~Roberts, Nucl. Phys. {\bf B455} (1995) 83;\\
P.M.~Ferreira, I.~Jack and D.R.T.~Jones, Phys. Lett. {\bf 357}
(1995) 359;\\ S.A.~Abel and B.C.~Allanach, Phys. Lett. {\bf B415}
(1997) 371;\\ J.~Casas, J.~Espinosa and H.~Haber, Nucl. Phys. {\bf
B526} (1998) 3;\\ M.~Jur\v{c}i\v{s}in and D.I.~Kazakov, Mod. Phys.
Lett. {\bf A14}(1999) 671;\\ G.~Auberson, G.~Moultaka, Eur. Phys. J.
{\bf C12} (2000) 331;
\\S.~Codoban, M.~Jur\v{c}i\v{s}in and
D.I.~Kazakov, Phys. Lett. {\bf B477} (2000) 223;\\
S.~Codoban, D.I.~Kazakov, Eur. Phys. J. {\bf C13} (2000) 671;\\
D.I.~Kazakov, G.~Moultaka, Nucl.Phys. {\bf B577} (2000) 121;\\
S.A.~Abel, B.C.~Allanach, JHEP 0007 (2000) 037;\\
I.~Jack, D.R.T.~Jones, Phys.Rev. {\bf D61} (2000) 095002;\\
C.-S.~Huang, L.~Wei, Q.-S.~Yan, S.-H.~Zhu, J.Phys. {\bf G27} (2001)
833;\\
Y.~Mambrini, G.~Moultaka, Phys. Rev. {\bf D65} (2002) 115011;\\
J.~Ferrardis, Phys.Rev. {\bf D68} (2003) 015001.
\bibitem{ja2}
G.K.~Yeghiyan, M.~Jur\v{c}i\v{s}in and D.I.~Kazakov, Mod. Phys.
Lett. {\bf A14}(1999) 601.
\bibitem{ja3}
M.~Jur\v{c}i\v{s}in, Proc. of Hadron Structure 2000, Stara Lesna,
Slovakia, October 2000, 326.
\bibitem{nmssm}
Y.~Mambrini, G.~Moultaka, M.~Rausch de Traubenberg, Nucl. Phys.
{B609} (2001) 83;\\
R.B.~Nevzorov, M.A.~Trusov, Phys. Atom. Nucl. {\bf 64} (2001) 1299;\\
R.B.~Nevzorov, M.A.~Trusov, Phys. Atom. Nucl. {\bf 64} (2001) 1513;\\
R.B.~Nevzorov, M.A.~Trusov, Phys. Atom. Nucl. {\bf 65} (2002) 335.
\bibitem{igo}
P.~Igo-Kemenes, Plenary talk ICHEP 2000, Osaka, Japan, July 2000.
\bibitem{feng}
J.L.~Feng, K.T.~Matchev and T.~Moroi, Phys. Rev. Lett. {\bf 84}
(2000) 2322;\\
J.L.~Feng, K.T.~Matchev and T.~Moroi, Phys. Rev. {\bf
D61} (2000) 075005.
\bibitem{focuss}
J.L.~Feng, K.T. Matchev, F.Wilczek, Phys. Lett. {\bf B482} (2000)
388;\\
J.L.~Feng, and  K.T. Matchev, Phys. Rev. {\bf D63} (2001) 095003;\\
J.L.~Feng, F.Wilczek, Phys. Lett. {\bf B631} (2005) 170.
\bibitem{rev0}
The CDF Collaboration, the D0 Collaboration, and the Tevatron
Electroweak Working Group, hep-ex/0507091.
\bibitem{davies}
C.T.H.~Davies, et al., Phys. Rev. {\bf D50} (1994) 6963.
\bibitem{rev1}
Review of Particle Properties, Phys. Rev. {\bf D50} (1994).
\bibitem{rev}
Review of Particle Properties, Eur. Phys. J. {\bf C3} (1998).
\bibitem{mtop1}
B. Schrempp, M. Wimmer, Prog. Part. Nucl. Phys. {\bf 37} (1996) 1.
\bibitem{mtop2}
D.M. Pierce, J.A. Bagger, K. Matchev and R. Zhang, Nucl.
Phys. {\bf B491} (1997) 3;\\
J.A. Bagger, K. Matchev and D.M. Pierce, Phys. Lett. {\bf B348}
(1995) 443;
\bibitem{Gl}
A.V. Gladyshev, D.I. Kazakov, W. de Boer, G. Burkart, R. Ehret,
Nucl. Phys. {\bf B498} (1997) 3;
\bibitem{cor}
M.  Carena, M.  Quiros, C.E.M.  Wagner, Nucl. Phys. {\bf B461} (1996) 407; \\
M. Carena, J.R. Espinosa, M. Quiros, C.E.M. Wagner,
Phys. Lett. {\bf B355} (1995) 209; \\
J. Ellis, G.L. Fogli, E. Lisi, Phys. Lett. {\bf B333} (1994) 118; \\
R. Hempfling, A. Hoang, Phys. Lett. {\bf B331} (1994) 99; \\
R. Hempfling, Phys. Rev. {\bf D49} (1994) 6168;\\
S. Heinemeyer, W. Hollik, G. Weiglein, Phys. Lett. {\bf B455} (1999) 179;\\
S. Heinemeyer, W. Hollik, G. Weiglein, Eur. Phys. J. {\bf C9}
(1999) 343.


\end{thebibliography}
\end{document}